\pdfoutput=1

\documentclass[11pt,twoside,a4paper,cmspaper,final,collab]{cms-tdr}
\def\svnVersion{461367P}\def\cmsCernNoTag{CERN-EP-2017-174}\def\cmsCernDate{\today}\def\cmsMessage{Published in Physical Review Letters as \href{http://dx.doi.org/10.1103/PhysRevLett.120.202301}{\doi{10.1103/PhysRevLett.120.202301.}}}

\begin{document}\cmsNoteHeader{HIN-16-007}

\hyphenation{had-ron-i-za-tion}
\hyphenation{cal-or-i-me-ter}
\hyphenation{de-vices}
\RCS$Revision: 461306 $
\RCS$HeadURL: svn+ssh://svn.cern.ch/reps/tdr2/papers/HIN-16-007/trunk/HIN-16-007.tex $
\RCS$Id: HIN-16-007.tex 461306 2018-05-23 00:26:29Z jisun $
\newlength\cmsFigOneWidth
\ifthenelse{\boolean{cms@external}}{\setlength\cmsFigOneWidth{0.90\columnwidth}}{\setlength\cmsFigOneWidth{0.5\textwidth}}
\newlength\cmsFigTwoWidth
\ifthenelse{\boolean{cms@external}}{\setlength\cmsFigTwoWidth{0.90\textwidth}}{\setlength\cmsFigTwoWidth{0.90\textwidth}}
\ifthenelse{\boolean{cms@external}}{\providecommand{\cmsLeft}{top\xspace}}{\providecommand{\cmsLeft}{left\xspace}}
\ifthenelse{\boolean{cms@external}}{\providecommand{\cmsRight}{bottom\xspace}}{\providecommand{\cmsRight}{right\xspace}}
\cmsNoteHeader{HIN-16-007}
\title{Measurement of prompt \texorpdfstring{$\PDz$ }{D0} meson azimuthal anisotropy in PbPb collisions at \texorpdfstring{$\sqrtsNNdef = 5.02\TeV $}{sqrt(s[NN]) = 5.02 TeV}}

\ifthenelse{\boolean{cms@external}}
{\newcommand{\sqrtsNNdef}{\ensuremath{\sqrt{\smash[b]{s_{_{N\!N}}}}}\xspace}}
{\newcommand{\sqrtsNNdef}{\ensuremath{\sqrt{\smash[b]{s_{_\mathrm{NN}}}}}\xspace}}
\newcommand{\Dzerodecay}{\ensuremath{\PDz \to \PKm\PGpp}\xspace}

\date{\today}

\abstract{The prompt $\PDz$ meson azimuthal anisotropy coefficients, $v_{2}$ and $v_{3}$, are measured
at midrapidity ($\abs{y}<1.0$) in PbPb collisions at a center-of-mass energy $\sqrtsNNdef = 5.02\TeV$ per nucleon pair with
data collected by the CMS experiment.
The measurement is performed in the transverse momentum (\pt)
range of 1 to 40\GeVc, for central and midcentral collisions.
The $v_{2}$ coefficient is found to be positive throughout the \pt range studied.
The first measurement of the prompt $\PDz$ meson $v_{3}$ coefficient is performed, and values up to
0.07 are observed for \pt around 4\GeVc.
Compared to measurements of charged particles, a similar \pt dependence, but smaller magnitude
for $\pt < 6\GeVc$, is found for prompt $\PDz$ meson $v_{2}$ and $v_{3}$ coefficients.
The results are consistent with the presence of collective motion of charm quarks at low \pt
and a path length dependence of charm quark energy loss at high \pt,
thereby providing new constraints on the theoretical description of the
interactions between charm quarks and the quark-gluon plasma.
}

\hypersetup{%
pdfauthor={CMS Collaboration},
pdftitle={Measurement of prompt D0 meson azimuthal anisotropy in PbPb collisions at sqrt(s[NN]) = 5.02 TeV},%
pdfsubject={CMS},%
pdfkeywords={heavy-ion, flow, quark gluon plasma, D-meson, open heavy-flavour}}

\maketitle

The formation of a strongly-coupled quark-gluon plasma (QGP), a state of matter
comprising deconfined quarks and gluons and exhibiting near-perfect liquid behavior,
was established first in experiments performed at the
Relativistic Heavy Ion Collider (RHIC)~\cite{Adams:2005dq,Adcox:2004mh,Arsene:2004fa,Back:2004je}
and then later confirmed at the CERN Large Hadron Collider (LHC)~\cite{Muller:2012zq,Armesto:2015ioy}.
The azimuthal anisotropy of produced light flavor particles,
one of the key signatures for the QGP formation,
can be characterized by the Fourier coefficients $v_{n}$
in the azimuthal angle ($\phi$) distribution of the hadron
yield, $\rd N/\rd \phi \propto 1 + 2 \sum_n v_{n}\cos[n(\phi-\Psi_{n})]$,
where $\Psi_{n}$ is the azimuthal angle of the direction of the maximum particle
density of the $n$th harmonic in the transverse plane~\cite{Poskanzer:1998yz}.
Heavy quarks (charm and bottom) are primarily produced via initial hard scatterings
because of their large masses, and thus carry information about the early stages of the QGP~\cite{BraunMunzinger:2007tn,Liu:2012ax}.
Detailed measurements of the azimuthal anisotropy of the final-state
charm and bottom hadrons can supply crucial information for understanding the properties of the QGP medium
and the interactions between heavy quarks and the medium~\cite{Andronic:2015wma}.
At low transverse momentum (\pt),
the charm hadron $v_{n}$ coefficient can help quantify the extent to which charm quarks flow with the medium,
which is a good measure of their interaction strength.
The measurements can also help explore the coalescence production mechanism for
charm hadrons where charm quarks recombine with light quarks from the medium,
which could also lead to positive charm hadron $v_{n}$~\cite{Molnar:2003ff,Greco:2003xt}.
At high \pt, the charm hadron $v_{n}$ coefficient can constrain the path length dependence of charm quark
energy loss~\cite{Gyulassy:2000gk,Shuryak:2001me}, complementing the
measurement of the nuclear modification factor~\cite{Adamczyk:2014uip,Adam:2015sza,Sirunyan:2017xss}.

The charm hadron $v_{2}$ coefficient has been studied indirectly by measuring the $v_{2}$ of leptons from
heavy-flavor hadron decays~\cite{Adare:2006nq,Adare:2010de,Adamczyk:2014yew,Adam:2016ssk,Adam:2015pga}.
The $\PD$ meson $v_{2}$ coefficient, which can provide cleaner information on the interactions between charm quarks and the medium,
has also been measured~\cite{Adamczyk:2017xur,Abelev:2014ipa,Acharya:2017qps}.
The $\PDz$ meson $v_{2}$ results from STAR
suggest that the charm quarks have achieved local thermal equilibrium with the QGP medium in the hydrodynamic picture~\cite{Adamczyk:2017xur}.
The $\PD$ meson $v_{2}$ values measured by ALICE are similar to those of light hadrons~\cite{Abelev:2014ipa,Acharya:2017qps}.
These results indicate that low-\pt charm quarks take part in the collective motion of the system.
The $\PD$ meson $v_{3}$ coefficient, which is predicted to be more sensitive to
the interaction strength between charm quarks and the medium~\cite{Nahrgang:2014vza},
has not been measured previously.
In general, a precise measurement of the $\PD$ meson $v_{n}$ coefficient over a wide momentum range is expected to provide valuable insight into the
QGP properties and can further constrain theoretical models.

In this Letter, we report the measurements of the azimuthal anisotropy coefficients, $v_{2}$ and $v_{3}$, of prompt
$\PDz$ mesons in lead-lead (PbPb) collisions at a center-of-mass energy $\sqrtsNNdef = 5.02\TeV $ per nucleon pair
with the CMS experiment at the LHC.
The coefficients are determined at midrapidity ($\abs{y}<1.0$) over a wide range in \pt (1 to 40\GeVc)
using the scalar product (SP) method~\cite{Adler:2002pu,Luzum:2012da}.
Results are presented for the centrality (i.e. the degree of overlap of the two colliding nuclei) classes 0--10\%, 10--30\%, and 30--50\%,
where the centrality class of 0--10\% corresponds to the 10\% of collisions with the largest overlap of the two nuclei.

The central feature of the CMS apparatus is a superconducting solenoid of 6\unit{m} internal
diameter, providing a magnetic field of 3.8\unit{T}. Within the solenoid volume are a silicon pixel and strip tracker,
a lead tungstate crystal electromagnetic calorimeter, and a brass and scintillator hadron calorimeter,
each composed of a barrel and two endcap sections.
Iron and quartz-fiber Cherenkov hadron forward (HF) calorimeters cover the pseudorapidity range
$3.0 < \abs{\eta}< 5.2$ on either side of the interaction region.
The granularity of the HF towers is $\Delta\eta{\times}\Delta\phi = 0.175{\times}0.175$ radians, allowing an
accurate reconstruction of the heavy ion collision event planes.
The silicon tracker measures charged particles within the pseudorapidity range $\abs{\eta} < 2.5$.
Reconstructed tracks with $1 < \pt < 10\GeVc$ typically have resolutions of 1.5--3.0\% in \pt
and 25--90 (45--150)\mum in the transverse (longitudinal) impact parameter \cite{TRK-11-001}.
A more detailed description of the CMS detector, together with a definition of the coordinate system used and the
relevant kinematic variables, can be found in Ref.~\cite{bib_CMS}.

The PbPb data used in this analysis
are selected by a minimum bias trigger and a 30--100\% centrality trigger.
The collision centrality is determined from the transverse energy (\ET) deposited in both HF calorimeters.
The minimum bias trigger requires energy deposits in both HF
calorimeters above a predefined threshold of approximately 1\GeV.
Furthermore, to increase the data sample in the 30--50\% centrality range,
a dedicated trigger is used to select events in the 30--100\% centrality range.
In the offline analysis, an additional selection of hadronic collisions is applied by
requiring at least three towers in each of the HF detectors with energy deposits of greater than 3\GeV per tower.
Events are required to have at least one reconstructed primary vertex, formed
by two or more associated tracks and required to have a distance from the
nominal interaction region of less than 15\unit{cm} along the beam axis.
The numbers of events used in the 0--10\%, 10--30\%, and 30--50\% centrality ranges are
$32\times10^{6}$, $64\times10^{6}$ and $151\times10^{6}$, respectively.

The $\PDz$ mesons (including both the $\PDz$ and $\PADz$ states) are reconstructed through the hadronic decay
channel \Dzerodecay, which has a branching fraction of ($3.93\pm0.04$)\%~\cite{Olive:2016xmw}.
The $\PDz$ candidates are formed by combining pairs of oppositely-charged
tracks and requiring an invariant mass within a ${\pm}200\MeVcc$ window of the
nominal $\PDz$ mass of 1864.83\MeVcc~\cite{Olive:2016xmw}.
Tracks are required to pass kinematic selections of $\pt > 0.7\GeVc$ and $\abs{\eta}<$ 1.5,
and must satisfy high-purity track quality criteria~\cite{TRK-11-001} to reduce the fraction of misreconstructed tracks.
For each pair of selected tracks, two $\PDz$ candidates are considered
by assuming one of the tracks has the pion mass while the other track has the kaon mass, and vice versa.
Kinematic vertex fits~\cite{Forden:1985xa} are performed to reconstruct the secondary vertices of $\PDz$ candidates.
Several selections related to the topology of the decay are applied in order to reduce the combinatorial background.
In particular, the selections are applied to
the three-dimensional (3D) decay length significance ($L_{xyz}/\sigma(L_{xyz})$),
defined as the 3D distance between the secondary and primary vertices divided by its uncertainty;
the pointing angle ($\theta_{p}$),
defined as the angle between the total momentum vector of the two tracks and the
vector connecting the primary and secondary vertices;
the $\chi^2$ probability of the secondary vertex fit;
and the distance of the closest approach ($\mathrm{DCA}$) of the total momentum vector to the primary vertex.
The signal-to-background ratios are \pt dependent, thus
\pt-dependent selection criteria are applied to $L_{xyz}/\sigma(L_{xyz})$ and the vertex probability,
ranging from 6.0 to 3.0 and 0.25 to 0.05 for low to high \pt, respectively.
In the selection, $\theta_{p} < 0.12$ radians and ${\mathrm{DCA}} < 0.008$\unit{cm} is required.
The selection on $\mathrm{DCA}$ not only increases the signal significance, but also suppresses the fraction of
nonprompt $\PDz$ ($\PDz$ mesons from decays of \PQb hadrons) significantly, which reduces the systematic uncertainties from
the nonprompt $\PDz$ meson contribution, as discussed later.

The event plane angles corresponding to the $n$th harmonic can be expressed
in term of Q-vectors, $Q_{n} = \sum_{k=1}^M \omega_{k}\re^{in\phi_k}$,
where $M$ represents the subevent multiplicity, $\phi_k$ is the azimuthal angle of the $k$th
particle, and $\omega_{k}$ is a weighting factor.
In this analysis, event planes determined from the two HF calorimeters covering the range $3<\abs{\eta}<5$, and
from the tracker using tracks within $\abs{\eta}<0.75$ are used. For the HF (tracker) event planes, $M$ is the number of
towers (tracks), and $\omega_{k}$ is the \ET deposited in each HF tower (\pt of each track).
The Q-vector of each $\PDz$ candidate is defined as $Q_{n,\PDz} = \re^{in\phi}$,
where $\phi$ is the azimuthal angle of the $\PDz$ candidate.
In the SP method, $v_{n}$ coefficient can be expressed in terms of the Q-vectors as:
\begin{linenomath}
\begin{equation}
\label{eq:vnSP}
{v_n}\left\{ {{\mathrm{SP}}} \right\} = {\frac{\left\langle {{Q_{n,\PDz}}Q_{nA}^*} \right\rangle } {\sqrt {{\frac{\left\langle {Q_{nA}^{}Q_{nB}^*} \right\rangle \left\langle {Q_{nA}^{}Q_{nC}^*} \right\rangle } {\left\langle {Q_{nB}^{}Q_{nC}^*} \right\rangle }}} }},
\end{equation}
\end{linenomath}
where the subscripts $A$ and $B$ refer to the HF event planes, the subscript $C$ refers to the tracker event plane,
and the $\left\langle \right\rangle$ in denominator (numerator) indicates an average over all events (all $\PDz$ candidates).
The denominator of Eq.~(\ref{eq:vnSP}) corrects for the finite resolution of the event plane $A$.
To avoid few-particle correlations, such as those induced by high-$\pt$ dijets and particle decays,
the $\eta$ gap between $\PDz$ candidates and the correlated event plane $A$ is required
to be at least three units.
Thus, if the $\PDz$ candidate comes from the positive-$\eta$ side,
$Q_{nA}$ ($Q_{nB}$) is calculated using the negative-$\eta$ (positive-$\eta$) side of HF, and vice versa.
The real part is taken for all averages of $Q$-vector products.
To account for asymmetries that arise from acceptance
and other detector-related effects, the Q-vectors of event planes are recentered~\cite{Poskanzer:1998yz,Alt:2003ab}.
These corrections and their effects on the results are found to be negligible.

\begin{figure}[h]
\centering
\includegraphics[width=\cmsFigOneWidth]{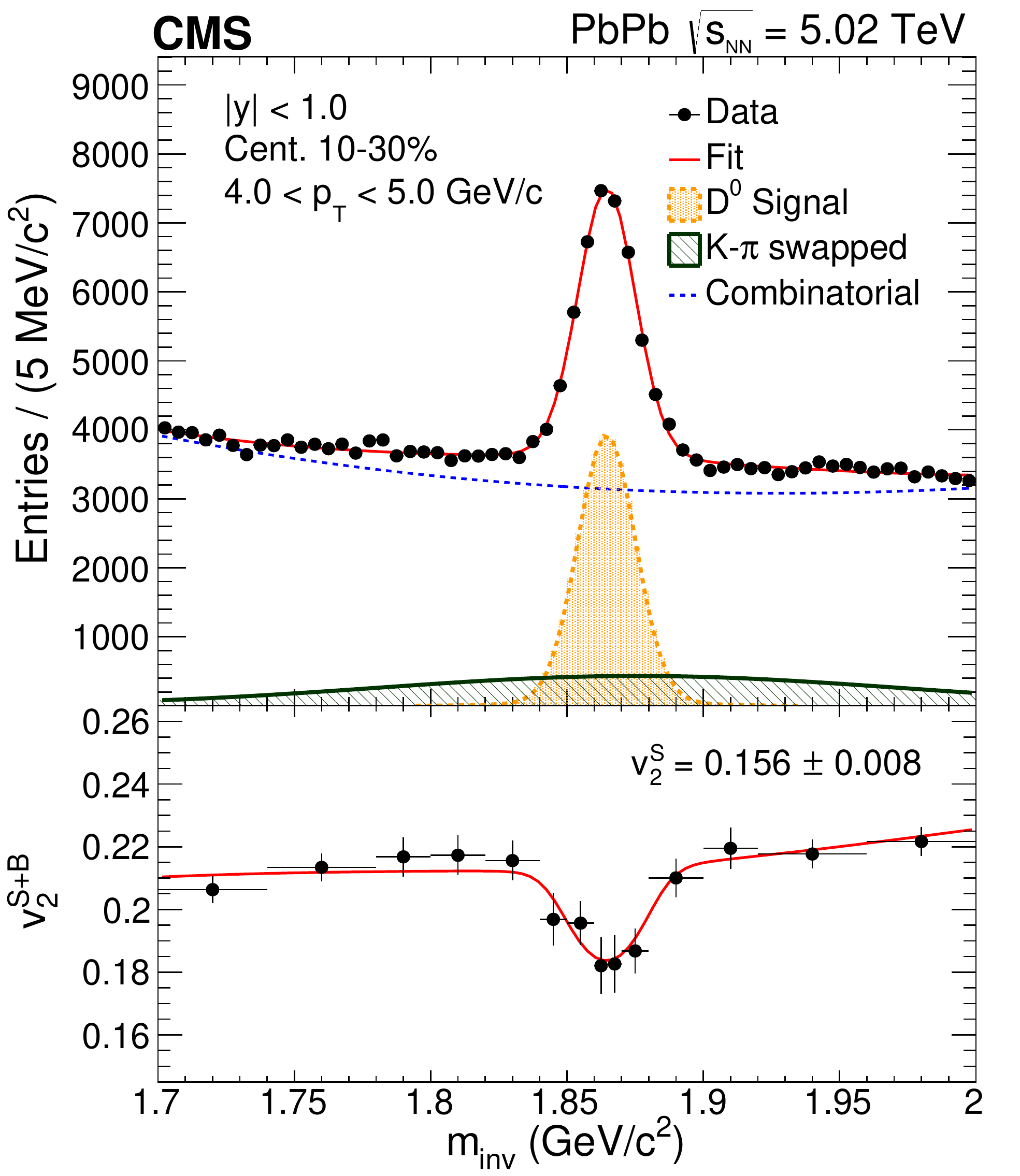}
\caption{Example of simultaneous fit to the invariant mass spectrum
and $v_{2}^{S+B}(m_\text{inv})$ in the \pt interval 4--5\GeVc for the centrality class 10--30\%.}
\label{fig:massv2fitcent10to30poly3bkg_floatwidth}
\end{figure}

To extract $v_{n}$ ($n = 2,\, 3$) values of the $\PDz$ signal ($v_{n}^{S}$),
a simultaneous fit to the invariant mass spectrum of $\PDz$ candidates and their $v_{n}$ distribution
as a function of the invariant mass ($v_{n}^{S+B}(m_\text{inv})$) is performed in each \pt interval.
The mass spectrum fit function is composed of three components: two Gaussian functions with the same mean but different
widths for the $\PDz$ signal ($S(m_\text{inv})$), an additional Gaussian function to describe the
invariant mass shape of $\PDz$ candidates with an incorrect mass assignment from the exchange of the pion and kaon designations ($SW(m_\text{inv})$),
and a third-order polynomial to model the combinatorial background ($B(m_\text{inv})$).
The width of $SW(m_\text{inv})$ is fixed according to
\textsc{pythia+hydjet} simulations, in which the $\PDz$ signal events from \textsc{pythia} 8.209~\cite{Sjostrand:2007gs,Khachatryan:2015pea}
are embedded into the minimum bias PbPb events from \textsc{hydjet} 1.9~\cite{Lokhtin:2005px}.
Furthermore, the ratio of the yields of $SW(m_\text{inv})$ and $S(m_\text{inv})$ is fixed to the value extracted from simulations.
The $v_{n}^{S+B}(m_\text{inv})$ distribution is fit with:
\begin{linenomath}
\begin{equation*}
v_{n}^{S+B}(m_\text{inv}) = \alpha(m_\text{inv})v_{n}^{S} + [1-\alpha(m_\text{inv})]v_{n}^{B}(m_\text{inv}),
\end{equation*}
\end{linenomath}
where
\begin{linenomath}
\begin{equation*}
\ifthenelse{\boolean{cms@external}}
{
\begin{split}
\alpha(m_\text{inv}) =& \\ [S(m_\text{inv}) & {+} SW(m_\text{inv})] / [S(m_\text{inv}) {+} SW(m_\text{inv}) {+} B(m_\text{inv}) ].
\end{split}
}
{
\alpha(m_\text{inv}) = [S(m_\text{inv}) + SW(m_\text{inv})] / [S(m_\text{inv}) + SW(m_\text{inv}) + B(m_\text{inv}) ].
}
\end{equation*}
\end{linenomath}
Here $v_{n}^{B}(m_\text{inv})$ is the $v_{n}$ value of background $\PDz$ candidates and
is modeled as a linear function of the invariant mass,
and $\alpha(m_\text{inv})$ is the $\PDz$ signal fraction as a function of the invariant mass.
The $\PK$-$\PGp$ swapped component is included in the signal fraction because
these candidates are from genuine $\PDz$ mesons and should have the same $v_{n}$ value as that of the true $\PDz$ signal.
Figure~\ref{fig:massv2fitcent10to30poly3bkg_floatwidth} shows an example of a simultaneous fit to the mass
spectrum and $v_{2}^{S+B}(m_\text{inv})$ in the \pt interval 4--5\GeVc for the centrality class 10--30\%.

The $\PDz$ signal in data is a mixture of prompt and nonprompt $\PDz$ components, thus
the $v_{n}^{S}$ is a combination of the $v_{n}$ coefficients of prompt $\PDz$ ($v_{n}^\text{prompt}$) and
nonprompt $\PDz$ ($v_{n}^\text{nonprompt}$) components,
\begin{linenomath}
\begin{equation*}
v_{n}^{S} = f_\text{prompt}v_{n}^\text{prompt} + (1-f_\text{prompt})v_{n}^\text{nonprompt},
\end{equation*}
\end{linenomath}
where $f_\text{prompt}$ is the fraction of prompt $\PDz$ mesons.
Besides the measurement of $v_{n}$ of $\PDz$ mesons with all analysis selections applied ($v_{n}^{S}$),
the $v_{n}$ of $\PDz$ mesons obtained by removing the $\mathrm{DCA} < 0.008\unit{cm}$ requirement ($v_{n,*}^{S}$)
and the corresponding prompt $\PDz$ fraction ($f_{\text{prompt},*}$) are also measured.
The prompt $\PDz$ fractions
are evaluated from data by fitting the $\mathrm{DCA}$ distribution
using the probability distribution functions for prompt and nonprompt $\PDz$ derived from the \textsc{pythia+hydjet} simulations.
The $\mathrm{DCA}$ distributions of the $\PDz$ signal in data are obtained with fits to mass spectra in bins of $\mathrm{DCA}$.
The discrimination between prompt and nonprompt $\PDz$ mesons lies mainly in the large $\mathrm{DCA}$ region, thus the
fit is performed on the entire range.
The $f_\text{prompt}$ and $f_{\text{prompt},*}$ are then evaluated from the fit.
It is found that the ${\mathrm{DCA}} < 0.008\unit{cm}$ requirement can suppress the fraction of nonprompt $\PDz$ mesons by approximately 50\%.
The $f_\text{prompt}$ ranges between 75 and 95\%, depending on \pt and centrality.
The $v_{n}^\text{prompt}$ can then be expressed as:
\begin{linenomath}
\begin{equation}
\label{eq:vnprompt}
v_{n}^\text{prompt} = v_{n}^{S} + \frac{1-f_\text{prompt}}{f_\text{prompt}-f_{\text{prompt},*}}(v_{n}^{S}-v_{n,*}^{S}).
\end{equation}
\end{linenomath}
The second term, $$\frac{1-f_\text{prompt}}{f_\text{prompt}-f_{\text{prompt},*}}(v_{n}^{S}-v_{n,*}^{S}),$$
is a correction factor to account for the remaining nonprompt $\PDz$ mesons after all analysis selections.
Taking the uncertainties in $f_\text{prompt}$ and $f_{\text{prompt},*}$ into account, the second term on the right of Eq.~(\ref{eq:vnprompt}) is
found to lie approximately between $-0.02$ and $+0.02$.
In this analysis, the $v_{n}^{S}$ values are kept as the central values of the measured prompt $\PDz$ meson $v_{n}$,
while the second term of Eq.~(\ref{eq:vnprompt}) is taken as a source of systematic uncertainty.

\begin{figure*}[t]
\centering
\includegraphics[width=\cmsFigTwoWidth]{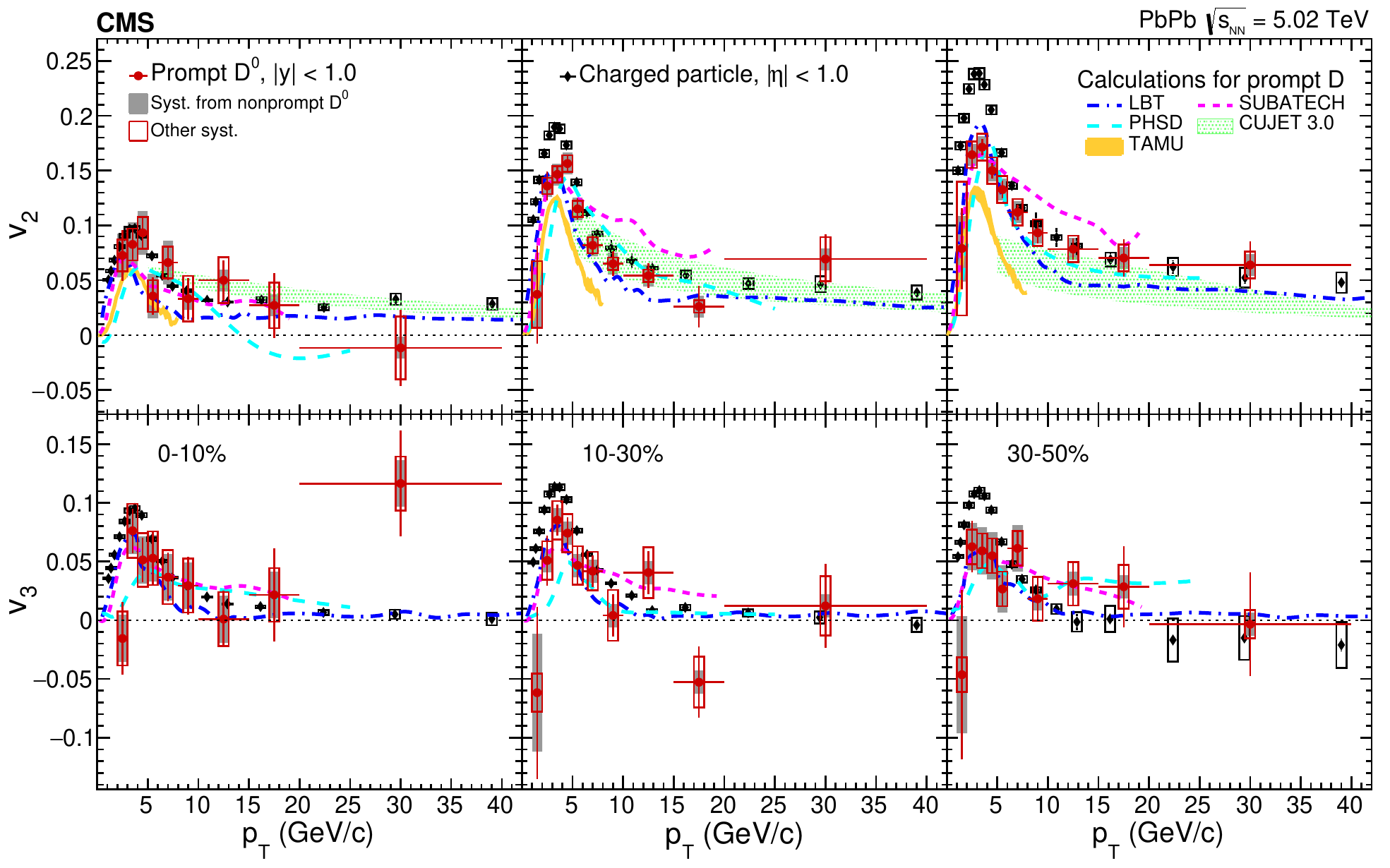}
\caption{Prompt $\PDz$ meson $v_{2}$ (upper) and $v_{3}$ (lower) coefficients at midrapidity ($\abs{y}<1.0$) for
the centrality classes 0--10\% (left), 10--30\% (middle), and 30--50\% (right).
The vertical bars represent statistical uncertainties, grey bands represent systematic uncertainties from nonprompt $\PDz$ mesons and
open boxes represent other systematic uncertainties.
The measured $v_{n}$ coefficient of charged particles at midpseudorapidity ($\abs{\eta}<1.0$)~\cite{Sirunyan:2017pan} and
theoretical calculations for prompt D
meson $v_{n}$ coefficient~\cite{PhysRevC.94.014909,Nahrgang:2014vza,Song:2015ykw,He:2014cla,Xu:2015bbz} are
also plotted for comparison.}
\label{fig:SPv2v3charged1theory1}
\end{figure*}

Apart from the systematic uncertainties from the remaining nonprompt $\PDz$ mesons discussed above,
other sources of systematic uncertainty in this analysis include the background mass
probability distribution function (PDF), the $\PDz$ meson yield correction (acceptance and efficiency),
the track selections, and the background $v_{n}$ PDF.
In this Letter, the quoted uncertainties in $v_{n}$ are absolute values.
The systematic uncertainties from the background mass PDF (0.001 for both $v_{2}$ and $v_{3}$)
are evaluated by the variations of $v_{n}$ while changing
the background mass PDF to a second-order polynomial or an exponential function.
Both the $\PDz$ meson yield correction, and the values of $v_{n}$ are functions of the $\PDz$ meson \pt,
so there will be systematic uncertainties arising from the correction.
To evaluate these uncertainties (0.002 to 0.003 for $v_{2}$ and 0.004 to 0.005 for $v_{3}$),
the yield correction is applied, and then $v_{n}$ values are extracted from the corrected
distributions and compared with the default $v_{n}$ values.
The track selections are also varied and systematic uncertainties
from track selections (0.005 to 0.02 for $v_{2}$ and 0.01 to 0.02 for $v_{3}$) are assigned based on the variations of $v_{n}$.
The systematic uncertainties from the background $v_{n}$ PDF (mostly 0.001 to 0.01 for $v_{2}$ and 0.005 to 0.015 for $v_{3}$)
are evaluated by changing $v_{n}^{B}(m_\text{inv})$ to a second-order polynomial function of the invariant mass.
The effects from few-particle correlations are also studied by varying the $\eta$ gap and are found to be negligible.

Figure~\ref{fig:SPv2v3charged1theory1} shows the prompt $\PDz$ meson $v_{2}$ (upper) and $v_{3}$ (lower)
coefficients at midrapidity ($\abs{y}<1.0$) for the centrality classes 0--10\% (left), 10--30\% (middle), and 30--50\% (right),
and compares them to those of charged particles (dominated by light flavor hadrons)
at midpseudorapidity ($\abs{\eta}<1.0$)~\cite{Sirunyan:2017pan}.
The $\PDz$ meson $v_{2}$ and $v_{3}$ coefficients increase with \pt to significant positive values in the low-\pt region,
and then decrease for higher \pt.
Compared to those of charged particles, the $\PDz$ meson $v_{2}$ and $v_{3}$ coefficients exhibit a similar \pt dependence.
As has been observed for charged particles, the $\PDz$ meson $v_{2}$ coefficient increases with decreasing
centrality in the 0--50\% centrality range,
while the $v_{3}$ coefficient shows little centrality dependence.
This is consistent with an increasing elliptical eccentricity with decreasing centrality~\cite{Chatrchyan:2012ta},
and an approximately constant triangularity stemming from geometry fluctuations~\cite{Chatrchyan:2013kba}.

For $\pt < 6\GeVc$, the magnitudes of $\PDz$ meson $v_{2}$ and $v_{3}$ coefficients are smaller than those for charged particles
in the centrality classes 10--30\% and 30--50\%.
Further study may determine whether it is a pure mass ordering or whether other effects,
such as the degree of charm quark thermalization, coalescence, and the path length dependence of energy loss, are at play.
The comparison between the $\PDz$ meson results and theoretical calculations in this low-\pt region (see discussion below) suggests
a collective motion of charm quarks.
For $\pt > 6\GeVc$, the $\PDz$ meson $v_{2}$ values remain positive, suggesting a path length dependence of the charm quark energy loss;
the $\PDz$ meson $v_{3}$ precision is limited by the available data.
The $\PDz$ meson $v_{2}$ values are consistent with those of charged particles,
suggesting that the path length dependence of charm quark energy loss is similar to that of light quarks.

Figure~\ref{fig:SPv2v3charged1theory1} also compares calculations from
theoretical models~\cite{PhysRevC.94.014909,Nahrgang:2014vza,Song:2015ykw,He:2014cla,Xu:2015bbz}
to the prompt $\PDz$ meson $v_{2}$ and $v_{3}$ experimental results.
The calculations from \textsc{lbt}~\cite{PhysRevC.94.014909}, \textsc{cujet} 3.0~\cite{Xu:2015bbz},
and \textsc{subatech}~\cite{Nahrgang:2014vza} include collisional and radiative energy losses, while those
from \textsc{tamu}~\cite{He:2014cla} and \textsc{phsd}~\cite{Song:2015ykw} include only collisional energy loss.
Initial-state fluctuations~\cite{Alver:2010gr} are included in the calculations from \textsc{lbt}, \textsc{subatech}, and \textsc{phsd};
thus calculations for the $v_{3}$ coefficient are only available from these three models.
For $ \pt < 6\GeVc$, \textsc{lbt}, \textsc{subatech}, \textsc{tamu}, and \textsc{phsd} can qualitatively describe the shapes of the measured $v_{2}$,
while the \textsc{tamu} model underestimates the $v_{2}$ values.
This may suggest that the heavy quark potential in the \textsc{tamu} model needs to be tuned~\cite{Liu:2015ypa} or
that the addition of radiative energy loss is needed.
The calculations from \textsc{lbt} and \textsc{subatech} are in reasonable agreement with the $v_{3}$ results,
while the \textsc{phsd} calculations are systematically below the measured $v_{3}$ for centrality class 10--30\%.
In the calculations from \textsc{lbt}, \textsc{subatech}, \textsc{tamu}, and \textsc{phsd},
the charm quarks have acquired significant elliptic and triangular flow
through the interactions with the medium constituents, and the coalescence mechanism is incorporated.
Without including the interactions between charm quarks and the medium, these models will
significantly underestimate the data~\cite{PhysRevC.94.014909,Nahrgang:2014vza,Song:2015ykw,He:2014cla}.
Thus, the fact that the calculated $v_{n}$ values are close to or even lower than the measured results
suggests that the charm quarks take part in the collective motion of the system.
Whether and how well the $\PDz$ anisotropy can be described by hydrodynamics and thermalization requires further investigation.
For $\pt > 6\GeVc$, \textsc{phsd} and \textsc{cujet} can generally describe the $v_{2}$ results.
\textsc{lbt} and \textsc{subatech} predict lower and higher $v_{2}$ values than in data, respectively,
indicating that improvements of the energy loss mechanisms in the two models are necessary.

In summary, measurements of prompt $\PDz$ meson azimuthal anisotropy coefficients, $v_{2}$ and $v_{3}$,
using the scalar product method in PbPb collisions at $\sqrtsNNdef = 5.02\TeV $ have been presented.
The $v_{2}$ values are found to be positive in the \pt range of 1 to 40\GeVc.
The $v_{3}$ coefficient is measured for the first time, and values up to 0.07 are observed for \pt around 4\GeVc.
The $v_{2}$ coefficient is observed to be centrality dependent, while the $v_{3}$ coefficient
shows little centrality dependence.
Compared with those of charged particles, the measured $\PDz$ meson $v_{2}$ and $v_{3}$
coefficients are found to be smaller
for $\pt < 6\GeVc$ but to have similar \pt dependence.
Through the comparison with theoretical calculations,
the $v_{2}$ and $v_{3}$ results at low \pt suggest that
the charm quarks take part in the collective motion of the system.
The $v_{2}$ values for $ \pt > 6\GeVc $, which are consistent with those of charged particles,
suggest that the path length dependence of charm quark energy loss is similar to that of light quarks.
The results provide new constraints on models of the interactions between charm quarks and the quark-gluon plasma,
and the charm quark energy loss mechanisms.

\begin{acknowledgments}
We congratulate our colleagues in the CERN accelerator departments for the excellent performance of the LHC and thank the technical and administrative staffs at CERN and at other CMS institutes for their contributions to the success of the CMS effort.
We thank E. Bratkovskaya, S. Cao, M. He, J. Liao, M. Nahrgang, R. Rapp, T. Song, and J. Xu for the inputs in comparing our measurements with their calculations.
In addition, we gratefully acknowledge the computing centers and personnel of the Worldwide LHC Computing Grid for delivering so effectively the computing infrastructure essential to our analyses. Finally, we acknowledge the enduring support for the construction and operation of the LHC and the CMS detector provided by the following funding agencies: BMWFW and FWF (Austria); FNRS and FWO (Belgium); CNPq, CAPES, FAPERJ, and FAPESP (Brazil); MES (Bulgaria); CERN; CAS, MoST, and NSFC (China); COLCIENCIAS (Colombia); MSES and CSF (Croatia); RPF (Cyprus); SENESCYT (Ecuador); MoER, ERC IUT, and ERDF (Estonia); Academy of Finland, MEC, and HIP (Finland); CEA and CNRS/IN2P3 (France); BMBF, DFG, and HGF (Germany); GSRT (Greece); OTKA and NIH (Hungary); DAE and DST (India); IPM (Iran); SFI (Ireland); INFN (Italy); MSIP and NRF (Republic of Korea); LAS (Lithuania); MOE and UM (Malaysia); BUAP, CINVESTAV, CONACYT, LNS, SEP, and UASLP-FAI (Mexico); MBIE (New Zealand); PAEC (Pakistan); MSHE and NSC (Poland); FCT (Portugal); JINR (Dubna); MON, RosAtom, RAS, RFBR and RAEP (Russia); MESTD (Serbia); SEIDI, CPAN, PCTI and FEDER (Spain); Swiss Funding Agencies (Switzerland); MST (Taipei); ThEPCenter, IPST, STAR, and NSTDA (Thailand); TUBITAK and TAEK (Turkey); NASU and SFFR (Ukraine); STFC (United Kingdom); DOE and NSF (USA).
\end{acknowledgments}

\bibliography{auto_generated}

\cleardoublepage \appendix\section{The CMS Collaboration \label{app:collab}}\begin{sloppypar}\hyphenpenalty=5000\widowpenalty=500\clubpenalty=5000\textbf{Yerevan Physics Institute,  Yerevan,  Armenia}\\*[0pt]
A.M.~Sirunyan, A.~Tumasyan
\vskip\cmsinstskip
\textbf{Institut f\"{u}r Hochenergiephysik,  Wien,  Austria}\\*[0pt]
W.~Adam, F.~Ambrogi, E.~Asilar, T.~Bergauer, J.~Brandstetter, E.~Brondolin, M.~Dragicevic, J.~Er\"{o}, M.~Flechl, M.~Friedl, R.~Fr\"{u}hwirth\cmsAuthorMark{1}, V.M.~Ghete, J.~Grossmann, J.~Hrubec, M.~Jeitler\cmsAuthorMark{1}, A.~K\"{o}nig, N.~Krammer, I.~Kr\"{a}tschmer, D.~Liko, T.~Madlener, I.~Mikulec, E.~Pree, D.~Rabady, N.~Rad, H.~Rohringer, J.~Schieck\cmsAuthorMark{1}, R.~Sch\"{o}fbeck, M.~Spanring, D.~Spitzbart, W.~Waltenberger, J.~Wittmann, C.-E.~Wulz\cmsAuthorMark{1}, M.~Zarucki
% ...
\vskip\cmsinstskip
% ...
1:~~Also at Vienna University of Technology, Vienna, Austria\\
% ...

\end{sloppypar}
\end{document}